\documentclass[prl,letter,twocolumn,epsf]{revtex4}
\usepackage{epsfig}
\usepackage{subfigure}
\usepackage{amsmath}
\usepackage{amssymb}
\usepackage{array}
\usepackage{pifont}
\usepackage{graphicx}

\def\ial35{Al$_{0.35}$Ga$_{0.65}$As}
\begin{document}

\title{Resonant Rayleigh Scattering in Ordered and
Intentionally Disordered Semiconductor Superlattices}

\author{V.\ Bellani, M.\ Amado$^*$}
\address{Dipartimento di Fisica ``A.Volta" and CNISM, Universit\`a degli Studi di Pavia, I-27100 Pavia, Italy}

\author{E.\ Diez}
\address{Departamento de F\'{\i}sica Fundamental, Universidad de Salamanca, E-37008 Salamanca, Spain}

\author{C.\ Koerdt$^{\dagger}$, M.\ Potemski}

\address{Grenoble High Magnetic Field Laboratory, CNRS, F-38042 Grenoble, France}

\author{R.\ Hey}
\address{Paul Drude Institut f\"ur Festk\"orperelektronik, D-10117 Berlin, Germany}

\date{\today}

\begin{abstract}

We report the experimental study of resonant Rayleigh scattering
in GaAs-AlGaAs superlattices with ordered and intentionally
disordered potential profiles (correlated and uncorrelated) in the
growth direction $z$. We show that the intentional disorder along $z$ modify
markedly the energy dispersion of the dephasing rates of the excitons. The
application of an external magnetic field in the same direction allows the
continuous tuning of the in plane exciton localization and to study
the interplay between the in plane and vertical disorder.

\end{abstract}


\maketitle


\narrowtext

\section{Introduction}

In recent years, a large number of studies has been devoted to
explore the general properties of disordered systems through the
investigation of the optical response and transport properties of
semiconductor nanostructures.
An important topic in the physics of
disordered systems is to understand the influence of an intentionally
disordered potential profile along the growth axis $z$ in semiconductor
superlattices (SL's)  on quantum interference and carrier localization~\cite{Chomette,Vittorio1,Yuri1}.
SL's are excellent candidates to perform such a
study since the advances achieved in molecular beam epitaxy (MBE)
allow the fabrication SL's tailored with desired conduction- and
valence-band profiles. In this way it is possible to design a great variety of one dimensional
potentials with intentional disordered in the growth direction $z$ (vertical disorder)
which coexists with the in plane disorder due to unintentional growth islands. The global result is an anisotropic disorder which effect on the quantum interference has become an interesting subject of recent experimental studies. As an example, magneto-transport measurements in doped semiconductor superlattices recently allowed the study of the effects of the anisotropy on the phase-breaking time of the electrons~\cite{Yuri1}.

In the present work we measure simultaneously the continuous wave resonant Rayleigh scattering (RRS), the photoluminescence(PL) and photoluminescence excitation (PLE) spectra of semiconductor SL's in a magnetic field to study the influence of the vertically correlated and uncorrelated anisotropic disorder on the coherence relaxation and exciton localization. We use this technique since RRS has access to information on the effective localization and on the spectral dispersion of the dephasing~\cite{Savona3} while
the photoluminescence(PL) and photoluminescence excitation (PLE) spectra in magnetic field are valid techniques to study the exciton delocalization along $z$~\cite{Vittorio3}. At the same time the external magnetic field allows to continuously tune the in plane exciton localization, with respect to the vertical one.

RRS is the resonant light scattering into all directions due to the
localization of the exciton center of mass wave functions, and is proved
to be a powerful tool for the study of dephasing process mainly in
quantum wells
(QWs)~\cite{Hegarty1,Hegarty2,Cantarero,Gurioli1,Gurioli2,Garro,
Garro2,Savona,Savona2,Runge}, but also in
microcavities~\cite{Gurioli3}, quantum wires~\cite{Lagoudakis} and
in bulk semiconductors~\cite{Garro,Allen}.
Rayleigh scattering in semiconductors arises form imperfection
breaking the symmetry of translational invariance of the crystal
due to imperfection in the bulk or on the surface, or from
interfaces. However crystals of very high quality can be grown by
means of MBE, and in this crystals the defect scattering can be
extremely low. In such systems and in particularly in GaAs-AlGaAs
multiple QWs, Hegarty and coworkers for the fist time observed a
strong contribution to the Rayleigh scattering in the vicinity of
an optical resonance~\cite{Hegarty1,Hegarty2}. In their pioneering
experiment the authors observed that RRS was enhanced 200-fold
with respect to the non resonant surface and defect scattering.
This RRS arises from the spatial fluctuation in the local resonant
frequency which, because of the strong dispersion near resonance,
leads to corresponding fluctuations in the refractive index.
Hegarty et al. showed also that RRS scattering is a valid experimental
technique to study the localization/delocalization of  two
dimensional excitons in QWs~\cite{Hegarty2}.
Despite the analogy between photoluminescence (PL) and RRS, these
two phenomena have a completely different natures. PL is an
incoherent process due to the relaxation of a real exciton
population. However RRS derives from the coherent scattering of
light as a consequence of the relaxation of wave-vector in a real
solid and mainly originate from localized exciton states~\cite{Garro,Gurioli1,
Savona2}.

Experimental spectroscopic techniques like time resolved RRS and
RRS spectral speclke analysis have been recently used with
success in order to extract quantitative information on
structural disorder from RRS (see, for instance Ref.~\cite{Savona3}
and references therein). In particular  the measured spectrally and time resolved
RRS spectra have been analyzed with accurate models which calculate the
RRS taking into account all the important features of semiconductor heterointerface
fabrication.
Specifically Savona and coworkers used either a time propagation of
the excitonic polarization with a
phenomenological dephasing or a full excitonic eigenstate model including
microscopically calculated dephasing rates and
phonon scattering rates~\cite{Savona3}.
The experimental spectra well agree with the calculations demonstrating the
role of many elements of the disorder affect the RRS~\cite{Savona3,Koche}.
Between these elements we have the interface structure, correlations and anisotropy.

\section{Experimental}

In this letter we study the RRS, PL and PLE in \ial35/GaAs ordered,
intentionally disordered and correlated disordered  SL in an
external magnetic field up to 14 T.
The excitation light was supplied by a Ti-sapphire laser pumped by an
Ar$^+$-ion laser. The excitation laser light was polarized $\sigma^+$
and the component $\sigma^+$ of the scattered and emitted light
has been detected.
The samples are three $n^+$-$i$-$n^+$ heterostructures with a SL grown
in the intrinsic region, grown by MBE. All the three samples have a SL of 200 periods and
\ial35 barriers with
width  of 3.2 nm. In the ordered SL (OSL) all the 200
wells are identical with well width  3.2 nm (hereafter referred to
as A wells). In the random  SL (RSL), it contains 142 A wells and
58 wells of thickness 2.6 nm (hereafter referred to as B wells)
and this B wells appears randomly. The so-called random dimer SL
(DSL) is identical to the RSL with the additional constraint that
the B wells appear only in pairs. In the latter sample the disorder
exhibit the desired short-range spatial correlations.
Additional details on the samples can be found in previous
publications~\cite{Vittorio1,Vittorio2}. Our previous theoretical
and experimental results~\cite{Diez,Vittorio1,Vittorio2} showed
that whereas in the RSL the states are localized,  the DSL
supports a narrow band of critical (non-Bloch like) extended
states, while the  OSL supports a wide band of Bloch extended
states. Therefore in the OSL the carrier are mobiles either in the
growth direction and in the plane. Instead for RSL and DSL we
found that the mobility is intimately related to the disorder and
that the exciton dynamics is affected by the presence of
correlations that  completely modify the localization of
the electronic states in the growth direction~\cite{Diez,Vittorio1}.

\section{Resonant Rayleigh scattering and Photoluminescence excitation}

In Fig. \ref{fig1} we report a series of emission spectra
(luminescence plus scattered light) measures on the OSL (a), DSL
(b) and RSL (c) respectively, scanning the energy of the
excitation laser light, for several hundreds different values of
the excitation energy  for polarized excitation and detection.
These emission spectra have been plotted using
small dots, and two main structures can be identified, one with
high density of points corresponding to PL  and the other with
less dense point corresponding to RRS. For clarity, in the insert
of the upper panel of Fig. \ref{fig1}(a) we show only 19 of these
spectra (for which the excitation energy differs of 0.8 meV from
one spectrum to the other). When the laser energy is set off
resonance we observe a weak Rayleigh scattering due to defect and
imperfections. Instead when the laser energy is set close to the
resonance the Rayleigh scattering increases dramatically and
reaches a peak some tenths times the off-resonance defect
scattering. As far as we know this is the first report of RRS in
semiconductor superlattices.
In the same figure we include the photoluminescence
excitation spectra (PLE) measured detecting at the energy of the
maximum of the PL peak. The PLE spectrum of the OSL (Fig.
\ref{fig1}(a)) shows a main peak due to the heavy hole exciton
(HH) and a smaller one due to the light hole one (LH). This
attribution of the LH transition is in agreement with the recent work of
Gerlovin et al. who measured PLE spectra in SL samples with
similar  well and barrier widths~\cite{Gerlovin}.
As regards the DSL, its PLE spectrum (Fig. \ref{fig1}(b))
shows of a main peak at 1.698 eV due to the transition between the
valence and conduction minibands which are present due to the
dimer type correlation of the disorder given by the paired 2.6 nm
wells. The spectrum present also a wider structure at 1.707 eV
which is due to excitonic transitions in the 3.2 nm isolated wells. These
two structure are observed also in the PL spectrum and in the RRS
since the latter is a scattering phenomenon resonant with the PL.
Finally for the RSL
(Fig. \ref{fig1}(c)) the PLE and the RRS spectrum of this sample
at zero magnetic field shows a main structure with a high energy
exponential tail typical of the disordered systems.

As regards the analysis of the RRS, Hegarty et al.~\cite{Hegarty1}
in their pioneering experimental study in semiconductor QWs
interpreted the experiments with the presence of a mobility edge
to explain the dispersion of the dephasing time.
Their work was described theoretically by
Takagahara who studied in detail the link between the
energy dependence of the homogeneous linewidth and
the dispersion of the dephasing time~\cite{Taka}.
More recent theories and experiments show that the
RRS can be interpreted without the need of introducing a
mobility edge~\cite{Savona3,Erland}.
In particular Savona et al. developped very accurate
theoretical calculations of the RRS using exiton
eigenstates and microscopically calculated dephasing rates due
to radiative decay and phonon scattering, which are
in very good agreement with the experiments~\cite{Savona3}.

The observation of a shift between RRS and PLE in semiconductor
QWs for the HH
transition was interpreted as a consequence of the inhomogeneous broadening
of the excitonic transition~\cite{Hegarty1,Hegarty2,Cantarero,Gurioli1,Gurioli2,Garro}.
In contrast in bulk GaAs and in GaAs-AlGaAs QWs
intentionally grown with high amount of interfacial disorder, the
shift of the RRS with respect to the PLE was found to be negligible~\cite{Garro}.
Accordingly to the description of Hegarty et al.~\cite{Hegarty1} the
shift is due to a rapid increase of the homogeneous linewidth
$\Gamma_h$. Instead the
absence of a shift between RRS and PLE can be interpreted as a
constant value for $\Gamma_h$ either in the case of localized or propagating
excitons.
With respect to the origin of the shift between RRS and PLE in semiconductor
QW's some authors authors ascribed it to the
presence of a mobility edge between propagating and localized
states~\cite{Hegarty1,Hegarty2,Gurioli1}.
However, as we mentioned previously, more recent experimental and theoretical
works show that the observed energy dispersion of the dephasing times
can be explained without the need of a mobility edge~\cite{Savona3,Erland}.

The three SL's that we study in this work have been grown in the same identical condition one after the other, and thus their in-plane disorder can be assumed to be same.
The aim of our work is to show that the RRS and PLE spectra are markedly different
in the thee SL' and their behaviour can be justified on the base of the different exciton
localization along $z$ due to the specific SL potential profile.
To this respect, the complete theoretical analysis of the different spectra that we
observe in the three SL's is a difficult task since requires accurate models~\cite{Savona3}
and goes beyond the aim of the present work.

The main PLE peak at zero magnetic field has a different shape and
width in the three SL's, due to their different electronic
structure. In particular it is wider and not symmetric in the RSL
and DSL, due to their intentional disorder (see Fig. \ref{fig1}).
This make difficult to determine the exact position of the PLE
peak in the RSL and DSL and therefore to compare the shift between
PLE and RRS in the three samples. However the position of the PLE
edge, taken as the energy at which the PLE low energy rapid rise
reaches the half value of its maximum, can be read with great
precision. Besides the energy region between the RRS and the PLE
edge is just the region where the dispersion of the dephasing
times takes place~\cite{Hegarty1,Taka}. Therefore we can read the
difference between the PLE edge and the RRS peak and compare it
for the three SL's. This energy difference is nearly 1 meV in the
OSL whereas is close to zero for the RSL and has an intermediate
value in the DSL.

This finding, i.e. the observation that the shift between RRS and PLE
absorption edge of the RSL is the smaller one, is compatible with the
previous interpretation  and suggest that in the OSL the presence of a
miniband in the $z$ direction allows the presence of exciton with a
great spectral dispersion of the dephasing rates. Within the seminal interpretation of Hegarty et al.~\cite{Hegarty1,Hegarty2,Gurioli1} this corresponds
to the presence of a mobility edge even if, as motioned previously, the latest
accurate theoretical models shows that the experimental RRS data and the
relevant dispersion of the dephasing time can be explained on the base
of accurate models without using the mobility edge concept.

In the case of the RSL the strong fluctuation of the potential profile in the growth
direction due to the intentional uncorrelated  disorder, induces
the exciton localization along $z$. This strong localization of
the excitons in the RSL leads to a smaller shift between RRS and PLE edge then in the
OSL. In the case of the DSL the presence, due to the correlation of the disorder, of a
narrow band of extended states ~\cite{Diez,Vittorio1,Vittorio2} increases
the exciton mobility along $z$ (with respect to that of the RSL) and
increases the energy dispersion of the dephasing time. This  gives a shift greater than
the one of the RSL, but in any case smaller that than in the case of the OSL which is the SL with the greater mobility along $z$.
The DSL  presents the unique characteristic of having
simultaneously in the same direction $z$ a potential profile
strongly disordered together with a band of mobile states, given
by the correlations. The greater value of the shift for the DSL, with
respect to the shift of the RSL is a clear indication that in the DSL
the presence of this extended states in the growth direction
reduces the exciton localization along $z$ influencing the
spectral dispersion of the dephasing times.

In Fig. \ref{fig2} we reports the same measurements of Fig.
\ref{fig1} but in the presence of a 14 T magnetic field applied in
the growth direction and in Faraday configuration. We observe that
applying a magnetic field the PLE, PL and the RRS spectra moves
towards higher energies due to the diamagnetic shift of the
excitons. In the meantime the HH peak widens and splits several
sub-structures. Also in the presence of the magnetic field
the DSL has a value of the shift that remains comprised between
the values of the RSL and the OSL, being the shift of the RSL the
smaller one. This indicate that the reduction of the localization
given by the correlation of the disorder is a robust phenomenon.We
observed the the magnetic field broadens and splits the RRS
spectra in several sub-structure due to the excited states of the
exciton therefore not allowing a sufficiently accurate
determination of the RRS position and, as a consequence, of the
shift.

\section{Photoluminescence shift and line width}

In Fig. \ref{fig3} we report the diamagnetic shift of the energy
of the PL as a function of the applied magnetic field applied in
the growth direction of the SL's. The dashed line is a quadratic
fit for OSL. We can observe that the energy of the main PL peak of
the three samples shift to higher energy in proportion to the
square of the magnetic field, as expected for excitonic
transitions~\cite{Sakaki}.  The shift is nearly the same for the
three SL's and by fitting with a quadratic expression the data for
the OSL (with well and barrier width of 2.6 nm) we find the
diamagnetic shift to be 22 $\mu$eV/T$^2$. This value is very close
to the earlier published value of 21 $\mu$eV/T$^2$ measured in QW
with well width of  2.1 nm~\cite{Rogers}. Our finding that the
diamagnetic shift of a SL is close to that of a QW with similar
well width was also observed in the past by Birkedal et
al.~\cite{Birkedal}. In their work they observed that a SL with
well and barrier width of 8 nm and 5 nm had a shift of 26.4
$\mu$eV/T$^2$ close to the value of 29 $\mu$eV/T$^2$ measured on
QW with 7.5 nm well width~\cite{Rogers}. Moreover our smaller
value of the diamagnetic shift with respect to the one measure by
Birkedal et al.~\cite{Birkedal} is coherent with the observations
that the diamagnetic shift increases with the increase of the well
width~\cite{Sakaki,Rogers}.

We also report in the inset of the Fig. \ref{fig3}
the full width at half maximum (FWHM) of the main PL peak of the
three SL's as a function of the magnetic field.
Al the FWHM refers to the PL measured with an excitation laser energy
0.025 eV above the PL peak position.
The linewidth of
the PL of the OSL increases very slightly with the magnetic field,
whereas the behaviour is not monotonous in the other two SL's. The
increase of the FWHM observed in the OSL has a very similar
with the one of semiconductor alloys~\cite{Jones,Singh,Mena,Lee} (we did
not found in literature papers on the dependence of the line-width
with the magnetic field in the case of SL's). In semiconductor alloys the
dominant mechanism for the PL line broadening is the statistical
potential fluctuation caused by the short-range alloy disorder.
The application of a magnetic field shrinks the exciton wave function,
hence reducing its effective size, and as a consequence an increasing of the
linewidth~\cite{Jones,Singh,Mena,Lee}.
In our OSL the carriers are free in the $x-y$ plane
and also move quite easily in the miniband in the $z$-direction.
Therefore we could assume the main source of disorder in our case
the in plane disorder of the interfaces between wells and barriers.
The typical length scale of
the growth islands is generally smaller that the exciton radius,
and is comparable with that of the alloy disorder and this will
explain the slight increase of the PL linewidth of the OSL with
the field.

In the two disordered SL's, the RSL and DSL,
the linewith changes not monotonously with the magnetic field.
In these SL's the main source disordered
is the well width intentional variation in the z-direction, with
length-scale greater than the exciton radius which. The competitive presence
of these intentional disorder and of the in-plane one
gives this non monotonous behaviour.
As we mentioned
previously the three SL have been grown in identical conditions so
we can reasonably assume that the in plane disorder is in average
the same. The observation that at all fields the value linewidth of the
DSL is comprised between that of the OSL and that of the RSL indicate
that the correlation of the disorder effectively reduces the localization
along $z$ and supporting the previous analysis of the RRS
regarding the dispersion of the dephasing rates.

The full theoretical study of the shift between RRS and PLE and
the calculation of the spectral dispersions of the dephasing
require accurate models capable to take into account the exciton
eigenstates for the three SL's and the microscopically calculated
dephasing rates due to radiative decay and phonon scattering,
taking into account the presence of the external magnetic field.
Also the quantitative interpretation of the modification of the PL
line-width with the field requires a full quantum statistical
theory of the system~\cite{Lee}. However these theoretical
analysis goes beyond the aim of this work which aim is to activate
and stimulate further interest in the understanding of basic the
properties of the disordered systems at different dimensionality.


In summary, we studied for the first time the RRS in semiconductor
SL's. This experimental technique allows to investigate in details
the electron localization and transport properties of the
two-dimensional electron gases in the wells. We measured ordered
SL's and intentionally disordered SL's also with correlated
disorder and we tuned the degree of localization of the excitons
in the wells applying an external magnetic field. We evidence that
the electronic properties along the growth direction strongly
affect the mobility of the electron gases of the wells. The
presence of extended states in the direction of the superlattice
potential profile reduces the exciton localization in the wells
and the coherence relaxation even thought this potential is
completely disordered.

\section{Acknowledgements}

We thank  F.\ Dom\'{i}nguez-Adame and A.\ Krokhin for the
enlightening discussions. We acknowledge the support of the
European Commission from the 6$^{th}$ framework programme
``Transnational Access-Specific Support Action"
(RITA-CT-2003-505474). E. Diez acknowledges supports from MEC
(Ram\'on y Cajal, FIS2005-01375, FIS2006-00716) and JCyL
(SA007B05) while M. Amado acknowledges supports from MEC
(MAT2003-1533).

\newpage

$^*$Present address: GISC, Departamento de F\'{\i}sica de
Materiales, Universidad Complutense, E-28040 Madrid, Spain.

$^{\dagger}$Present address: Institute for Scientific Computing,
Forschungszentrum Karlsruhe, D-76344, Germany

\newpage


\begin{figure}
\caption{Scattered light spectra (small points) and PLE spectra
(dashed line)  of the Ordered SL (a), Random SL (b) and Random
dimer-SL (c) at $4\,$ K and zero external magnetic field
for $\sigma^+$ excitation and detection polarization. In the
insert of panel (a) we show only some of these spectra (for
which the excitation energy differs of 0.8 meV from one spectrum
to the other).} \label{fig1}
\end{figure}

\begin{figure}
\caption{ Scattered light spectra (small points) and PLE spectra
(dashed line)  of the Ordered SL (a), Random SL (b) and Random
dimer-SL (c) in an external magnetic field of 14 T for $\sigma^+$
excitation and detection polarization.} \label{fig2}
\end{figure}

\begin{figure}
\caption{Energy shift of the energy of the PL for the three SL's as a
function of the magnetic field. The dashed line is a quadratic fit
of the shift for the Ordered SL The insert of panel (c) shows  the
full width half maximum (FWHM) of the PL peak as a function of the
magnetic field.}\label{fig3}
\end{figure}

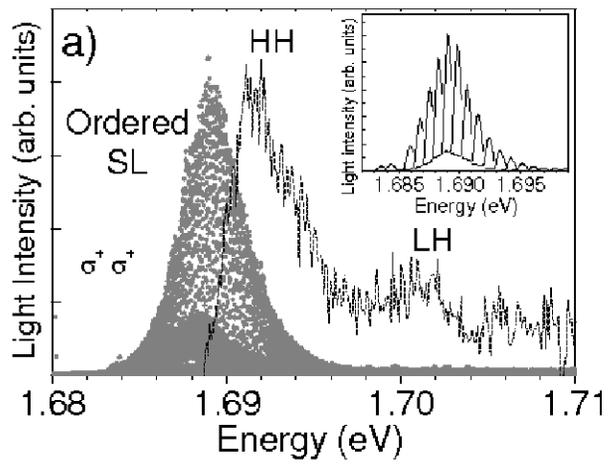
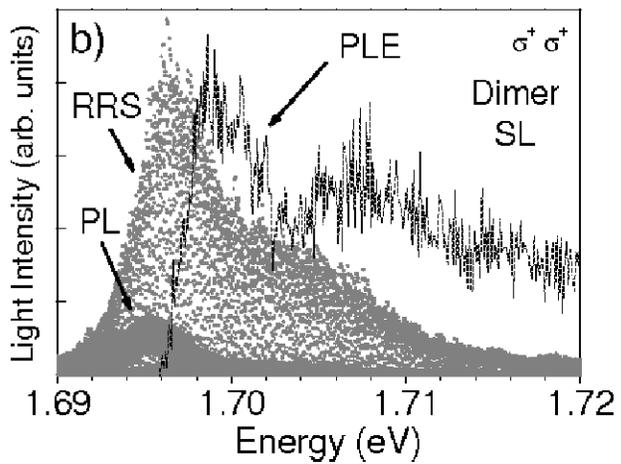
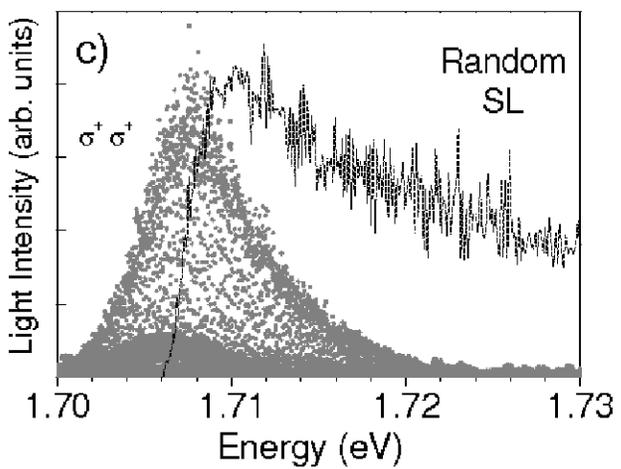
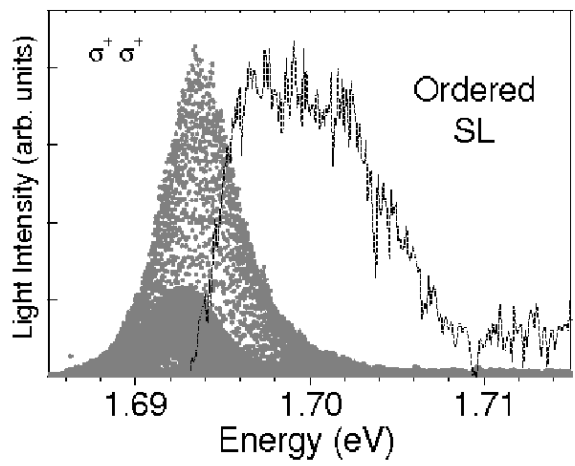
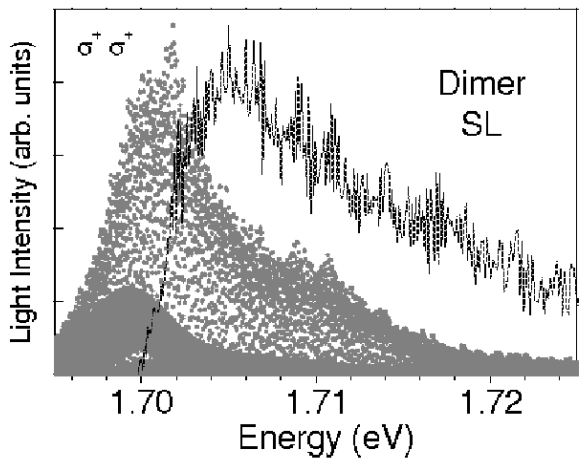
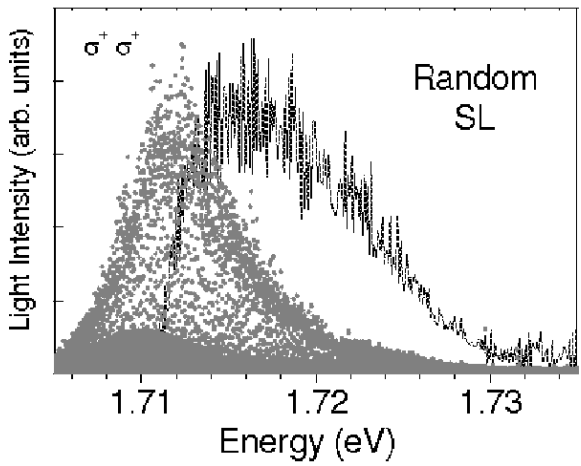
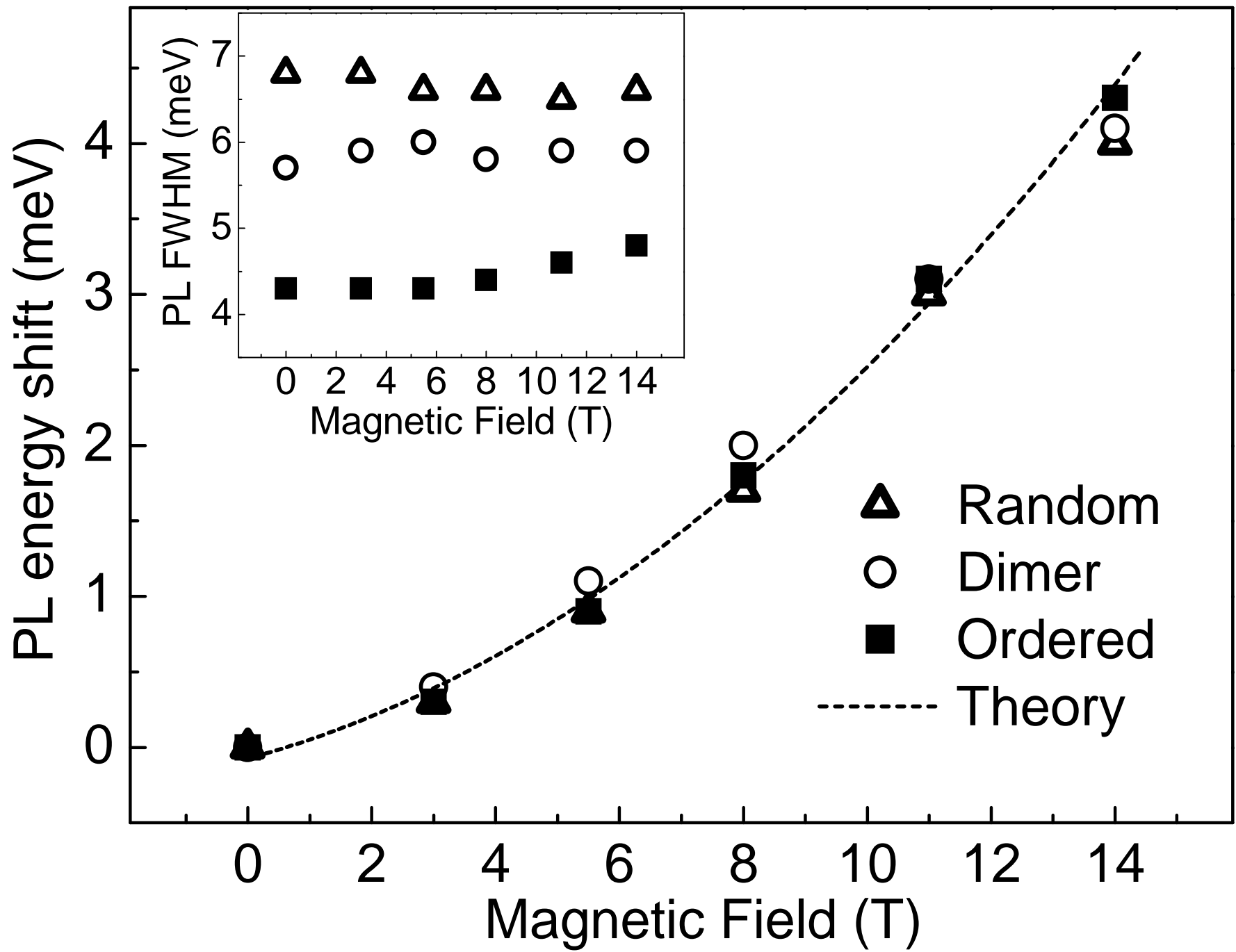
\end{document}